\documentclass[review]{elsarticle}

\usepackage{geometry}
\usepackage{numcompress}\biboptions{authoryear}

\graphicspath{{pic/}}
\date{}
\usepackage{booktabs}
\usepackage{siunitx}
\usepackage{amsmath}
\usepackage{subfigure}
\usepackage{float}
\usepackage[capitalize]{cleveref}
\usepackage{xcolor}
\usepackage{amsfonts}
\usepackage{bm}
\usepackage{diagbox}
\usepackage{url} 
\usepackage{graphicx}

\usepackage{caption}
\newcommand{\blueblack}{\color{black}}

\begin{document}

\begin{frontmatter}

\title{Flow-induced vibration of twin-pipe model with varying mass and damping: A study using virtual physical framework}

\author[mymainaddress,mysecondaryaddress,NTNUaddress]{Jiawei Shen}

\author[mymainaddress,mysecondaryaddress]{Shixiao Fu\corref{mycorrespondingauthor}}
\cortext[mycorrespondingauthor]{Corresponding author}
\ead{shixiao.fu@sjtu.edu.cn}
\author[mymainaddress,mysecondaryaddress]{Xuepeng Fu}

\author[NTNUaddress,NBUaddress]{Torgeir Moan}

\author[NTNUaddress]{Svein Sævik}

\address[mymainaddress]{State Key Laboratory of Ocean Engineering, Shanghai Jiao Tong University, Shanghai, 200240, China}
\address[mysecondaryaddress]{Institute of Polar and Ocean Technology, Institute of Marine Equipment, Shanghai Jiao Tong University, Shanghai, 200240, China}
\address[NTNUaddress]{Department of Marine Technology, Norwegian University of Science and Technology, Trondheim, 7491, Norway}
\address[NBUaddress]{Faculty of Maritime and Transportation, Ningbo University, Ningbo, China}

\begin{abstract}{\blueblack
Flow-induced vibration (FIV) commonly occurs in rigidly coupled twin-pipe structures. However, the limited understanding of their FIV responses and hydrodynamic features presents a major challenge to the development of reliable engineering designs. To bridge this gap, the present study systematically investigates the FIV characteristics of a rigidly coupled twin-pipe model with elastic support using a virtual physical framework (VPF), which enables flexible control of structural parameters during physical testing. 
A distinctive feature of twin-pipe structures is the presence of in-line hydrodynamic interactions and torsional moments arising from the rigid coupling. The in-line interaction is primarily compressive and becomes more pronounced as the mass ratio increases. The torsional moment coefficient exhibits a rise–fall trend with increasing reduced velocity \( U_R\) and stabilizes around 0.46 at low mass ratios.
In addition, an ``amplitude drop'' phenomenon is observed near \( U_R = 6 \), attributed to energy dissipation from the downstream pipe. The mass ratio significantly affects FIV amplitude, frequency, and hydrodynamic coefficients. As the mass ratio decreases, the synchronization region broadens and the hydrodynamic coefficients become more stable. At mass ratio of 1.0, a ``resonance forever'' behavior is observed. Damping primarily suppresses FIV amplitude, with minimal impact on dominant frequency and hydrodynamic coefficients. These findings provide valuable insights into twin-pipe FIV mechanisms and support a scientific basis for future structural design optimization.
}
\end{abstract}


\end{frontmatter}


\section{Introduction}

Twin-pipe structures are widely employed in offshore engineering applications, such as subsea pipelines, marine riser clusters, and submerged floating tunnels \citep{janocha2021flow, zhao2023experimental, deng2020experimental}. In these configurations, twin pipes are often rigidly connected to enhance economic efficiency and operational feasibility. Similarly, the submerged floating tunnel consists of two identical pipes that are rigidly connected by truss elements, ensuring structural integrity and stability \citep{deng2020experimentalD}.
These twin-pipe structures are susceptible to FIV, which arises from alternating vortex shedding and induces transverse oscillations on the structure \citep{zhao2023drag, fu2024vortex}. Strong flow interference in twin-pipe configurations alters vortex formation and wake patterns, resulting in a FIV behavior that differs significantly from those of an isolated pipe \citep{zhang2022wake, mysa2016origin}.
However, experimental data supporting the design of such structures remain limited. This limitation is particularly critical for submerged floating tunnels, where the hydrodynamic interaction between the two pipes imposes significant constraints on safe and efficient design.
Given these practical configurations and hydrodynamic complexities, investigating the FIV response and hydrodynamic characteristics of rigidly coupled tandem twin-pipe structures is essential for both scientific understanding and engineering application.

Elastically mounted rigid segments are commonly used as fundamental models in mechanistic studies of slender structures to explore fluid–structure interaction mechanisms at the element level \citep{williamson2004vortex}.  Most research on rigidly coupled tandem twin-pipe segments has primarily relied on numerical simulations. \cite{zhao2013flow} numerically investigated the effect of spacing ratio on the FIV response of rigidly coupled tandem twin pipes on a flexible mount at Re = 150, showing that smaller spacing ratios result in narrower lock-in regions, whereas larger spacing ratios lead to broader ones. \cite{zhu2023vortex} explored the influence of degrees of freedom (DoF) of rigidly coupled tandem twin pipes on FIV, highlighting its impact on transverse amplitude, lock-in region, and vortex interaction dynamics. While numerical methods provide valuable insights, experimental investigations are essential for a comprehensive understanding of the underlying mechanisms. \cite{shen2024experimental} experimentally examined the effect of spacing ratio $(G/D$) on the FIV response of rigidly coupled tandem twin pipes, revealing that $G/D = 1.5$ yielded the lowest response amplitude with a separated VIV-galloping feature, $G/D = 3$ showed amplitude modulation, and $G/D \geq 4$ resulted in similar response trends. \cite{deng2020experimentalD} experimentally studied the effects of spacing ratio and submergence on drag characteristics of rigidly coupled tandem twin pipes, and reported that smaller spacing ratios increase hydrodynamic interaction effects and amplify drag under FIV, while reduced submergence decreases both drag and FIV response. 

Despite extensive research on the effects of spacing ratio and degrees of freedom on FIV in rigidly coupled tandem twin pipes, the influence of mass and damping ratios remains insufficiently explored. These facts leave critical gaps in understanding the global FIV behavior of twin-pipe models to support practical design decisions. Research on the FIV of a single pipe has demonstrated that both parameters significantly affect vibration responses \citep{bahmani2010effects, govardhan2006defining, khalak1997investigation}. These findings highlight the need to study FIV of twin-pipe models with varying mass and damping ratios, both to improve hydrodynamic understanding and to provide information for engineering design. However, conducting twin-pipe FIV experiments with varying mass ratios introduces a significant challenge in mitigating the influence of inertial forces. For the pipe model used in this study, the hydrodynamic forces $F_h \sim \mathcal{O}(10^0)$ N are two orders of magnitude smaller than the inertial forces $F_i \sim \mathcal{O}(10^2)$ N, given a mass ratio $m^* = 10$.
This large difference in force magnitude makes it difficult for force sensors to achieve both sufficient range and high accuracy, which can lead to significant errors in hydrodynamic force measurements and affect the reliability of the hydrodynamic analysis.

To minimize the influence of inertial forces on hydrodynamic measurements as much as possible and to enable convenient and rapid adjustments of structural mass and damping, it is necessary to adopt an appropriate experimental technique that allows precise control and flexible parameter modification. Hybrid approaches integrating numerical simulations and physical model tests offer promising solutions for precise control of structural parameters in experimental studies \citep{ehlers2022committee}. \cite{hover1998forces} were pioneers in applying a hybrid method, called force-feedback control, to fluid-structure interaction experiments on a single pipe, using it to study free and forced vibrations in cross-flow (CF) direction. Building on this concept, \cite{mackowski2011developing} developed a cyber-physical fluid dynamics (CPFD) facility based on a discretized form of Newton’s law, which enabled straightforward force modification. However, time delays from noise filtering and motion execution remain a challenge, highlighting the need for improved delay compensation techniques. \cite{ren2024developing} established an enhanced virtual physical system (VPS) using a recursive Duhamel integral method (DIM) for real-time force-motion control, ensuring stability, precise parameter tuning, and effective delay compensation. While these advancements have significantly promoted the application of hybrid methods in FIV studies involving single-pipe models, their implementation in twin-pipe FIV experiments remains unexplored.

This study develops a VPF for rigidly coupled tandem twin-pipe models, enabling the investigation of FIV with different structural parameters. Based on this framework, systematic FIV experiments are conducted with varying mass and damping ratios. The content of this paper is organized as follows:
Section 1 introduces the background and fundamental methodology of the VPF method for twin-pipe FIV experiments.
Section 2 presents the experimental setup based on the VPF framework, including the test matrix and validation experiments to ensure reliability of VPF based setup. Section 3 provides detailed results and discussion, where the effects of mass and damping ratios on the FIV responses and hydrodynamic characteristics of the rigidly coupled tandem twin-pipe model are systematically analyzed. Special attention is given to in-line hydrodynamic interactions and torsional moments, which are of particular importance in the design of submerged floating tunnels.

\section{Methodology of VPF for twin-pipe model}\label{m}

\begin{figure}[htbp!]
	\centering
	\includegraphics[width=0.45\textwidth]{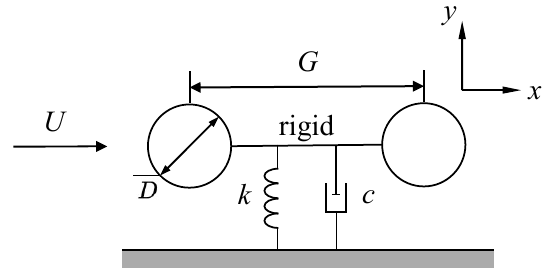}
	\caption{Schematic of the rigidly coupled tandem twin-pipe model, illustrating the definition of spacing ratio $G/D$, and coordinate system.}
	\label{geometry}
\end{figure}

This study develops a virtual physical framework (VPF) for a rigidly coupled tandem twin-pipe model with elastic support (hereinafter referred to as the twin-pipe model), enabling the investigation of its flow-induce vibration (FIV) with varying mass and damping ratios. \cref{geometry} illustrates the geometry of two-dimensional system with two rigidly connected pipes with elastic support, and the coordinate system. The spacing ratio is defined by \( G/D \), where \( G \) is the center-to-center distance between the two pipes and \( D \) is the pipe diameter. In this study, a spacing ratio of \( G/D = 2 \) is adopted. The investigation focuses on FIV in the cross-flow direction, along the \( y \)-axis, where the two pipes exhibit identical transverse displacement \( y(t) \) due to the rigid connection. Within VPF, a twin-pipe model is physically constructed and placed in a real uniform flow field generated by towing, while its elastic boundary conditions, damping, and mass are numerically simulated. This hybrid method allows for a systematic investigation of FIV responses and hydrodynamic features of a twin-pipe model with varying mass and damping ratios.

\begin{figure}[htbp!]
	\centering
	\includegraphics[width=0.7\textwidth]{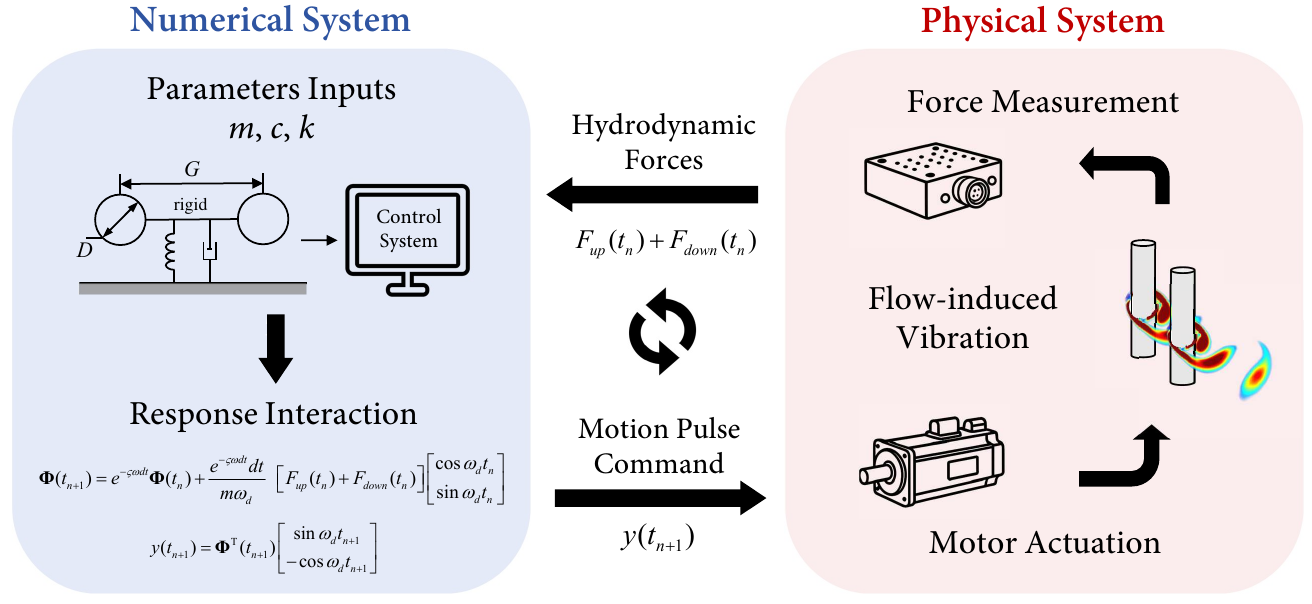}
	\caption{Schematic of the VPF for FIV experiments on a rigidly coupled tandem twin-pipe model. The framework integrates a numerical system, which computes structural responses and outputs motion commands, with a physical system that measures hydrodynamic forces and executes motor actuation, forming a closed-loop interaction.}
	\label{VFS}
\end{figure}

\cref{VFS} illustrates the schematic of the VPF for FIV experiments on twin-pipe models, where numerical and physical systems interact iteratively to capture fundamental physical processes. In VPF, multiple types of mass are defined and should be clearly distinguished. The virtual total mass $m$ represents the target total mass of the twin-pipe model in the simulation. The input mass $m_{\text{in}}$ is used in the numerical system and is specifically set to eliminate the effect of inertial forces, as will be described in detail later. The physical mass $m_p$ refers to the real mass of the carbon fiber twin-pipe model, including the upstream and downstream pipes, $m_{p,\text{up}}$ and $m_{p,\text{down}}$, which contribute directly to inertial forces in the measured hydrodynamic data. Additionally, the added mass of the twin-pipe model $m_a$ which arises from fluid acceleration effects, is included in the measured hydrodynamic forces and does not need to be explicitly modeled within the VPF.

The specific concept of the VPF is as follows: structural parameters of the twin-pipe model, including total mass ($m$), damping ($c$), and stiffness ($k$), are digitally set in the numerical system. At each time step, the flow-induced hydrodynamic forces in cross-flow direction are measured from the physical system using force sensors mounted at each end of the two pipes. The measured forces are then input into the numerical system, where a dynamic response solver calculates the displacement at a certain time step. Motion pulse commands are transmitted from the numerical system to the motion actuators within the physical system for precise displacement execution. Subsequently, updated hydrodynamic forces are measured and fed back into the numerical solver at each time step, establishing a closed-loop computational-physical interaction cycle. This iterative process operates with a minimal time increment of 0.001 s to mitigate discretization-induced errors, thereby enabling accurate simulations for the dynamic response of an elastically mounted structural system.

The dynamic response solution adopted in the numerical system of this study is the Duhamel Integral Method (DIM) in a time integral format \citep{arosio1984duhamel}.
This method is based on the dynamic response analysis method for impulsive loads, treating the complete load time history as a series of short impulses. Each impulse generates an impulse response. For linear elastic systems, the total response can be obtained by summing up all the differential responses generated by the load time history. For a rigidly coupled twin-pipe system with elastic boundaries, the motion of both pipes is synchronized and represented by \( y(t) \). The external hydrodynamic loads acting in the cross-flow direction on the upstream and downstream pipes, \( F_{\text{up}}(t) \) and \( F_{\text{down}}(t) \), collectively influence the dynamic response. The response \( y(t) \) under these loads can be calculated as follows:
\begin{equation}
y(t) = \frac{1}{m\omega_d} \int_{0}^{t} \left[ F_{\text{up}}(\tau) + F_{\text{down}}(\tau) \right] 
e^{-\zeta\omega (t-\tau)} \sin \omega_d (t-\tau) \, d\tau, \quad t \geq 0,
\label{eq:modified}
\end{equation}
where, $m$ is the total mass of the system, $\omega$ is the natural angular frequency of the system, $\zeta$ is the structural damping ratio, $\omega_d$ is the damped angular frequency of the system, calculated by $\omega_n \sqrt{1 - \zeta^2}$.

To facilitate the discrete implementation of the Duhamel integral method (DIM) and improve computational efficiency, \cref{eq:modified} is reformulated as a recursive algorithm. The displacement response \( y(t_n) \) at time step \( t_n \) is expressed as:
\begin{equation}
y(t_n) = \boldsymbol{\Phi}^\mathsf{T}(t_n)
\begin{bmatrix}
\sin \omega_d t_n \\
- \cos \omega_d t_n
\end{bmatrix},
\end{equation}
where \( \boldsymbol{\Phi}(t_n) = \begin{bmatrix} \alpha(t_n) & \beta(t_n) \end{bmatrix}^\mathsf{T} \) is the vector of temporary integral coefficients defined by:
\begin{equation}
\boldsymbol{\Phi}(t_n) =
\frac{e^{-\zeta \omega t_n}}{m \omega_d}
\int_0^{t_n}
\left[ F_{\text{up}}(\tau) + F_{\text{down}}(\tau) \right]
e^{\zeta \omega \tau}
\begin{bmatrix}
\cos \omega_d \tau \\
\sin \omega_d \tau
\end{bmatrix}
\, d\tau.
\label{eq:coefficients}
\end{equation}
At the next time step, two temporary integral coefficients $\alpha(t_{n+1})$, $\beta(t_{n+1})$ can be calculated as,
\begin{equation}
\boldsymbol{\Phi}(t_{n+1}) =
e^{-\zeta \omega \, dt} \, \boldsymbol{\Phi}(t_n)
+
\frac{e^{-\zeta \omega t_{n+1}}}{m \omega_d}
\int_{t_n}^{t_{n+1}}
\left[ F_{\text{up}}(\tau) + F_{\text{down}}(\tau) \right]
e^{\zeta \omega \tau}
\begin{bmatrix}
\cos \omega_d \tau \\
\sin \omega_d \tau
\end{bmatrix}
\, d\tau.
\end{equation}

In the VPF framework applied to the rigidly coupled twin-pipe system, the external forces acting on the upstream and downstream pipes, \( F_{\text{up}}(t_{n+1}) \) and \( F_{\text{down}}(t_{n+1}) \), are generally unknown at the future time step \( t_{n+1} \). A practical approach is to reduce the time step sufficiently small relative to the characteristic period of the external forces. Thus, it is assumed that the sum of the external forces remains approximately constant between consecutive time steps:
\begin{equation}
F_{\text{up}}(t_{n+1}) + F_{\text{down}}(t_{n+1}) \approx F_{\text{up}}(t_n) + F_{\text{down}}(t_n).
\label{eq:force_assumption}
\end{equation}
Under this assumption, the integral coefficients can be further expressed as follows:
\begin{equation}
\boldsymbol{\Phi}(t_{n+1}) =
e^{-\zeta \omega dt} \, \boldsymbol{\Phi}(t_n)
+
\frac{e^{-\zeta \omega dt} \, dt}{m \omega_d}
\left[ F_{\text{up}}(t_n) + F_{\text{down}}(t_n) \right]
\begin{bmatrix}
\cos \omega_d t_n \\
\sin \omega_d t_n
\end{bmatrix}
\end{equation}
Thus, the displacement at the next time step can be calculated as:
\begin{equation}
y(t_{n+1}) =
\boldsymbol{\Phi}^\mathsf{T}(t_{n+1})
\begin{bmatrix}
\sin \omega_d t_{n+1} \\
- \cos \omega_d t_{n+1}
\end{bmatrix}
\end{equation}
With these integral expressions established, the recursive implementation of the DIM within the VPF framework can now be formulated. This approach enables an efficient step wise determination of the next-step motion displacement instruction based on the current measured forces, without the need to integrate the entire force history over time.

In addition, it should be noted that the measured forces include the inertial forces resulting from the physical mass of each pipe, which can be expressed in vector form as:
\begin{equation}
\boldsymbol{\mathcal{F}}_m(t_n) =
\boldsymbol{\mathcal{F}}(t_n) - \boldsymbol{m}_p \, \ddot{y}(t_n),
\label{eq:force_total}
\end{equation}
where \( \boldsymbol{\mathcal{F}}_m(t_n) = \begin{bmatrix} F_{\text{up,m}}(t_n) & F_{\text{down,m}}(t_n) \end{bmatrix}^\mathsf{T} \) denotes the vector of measured forces,  
\( \boldsymbol{\mathcal{F}}(t_n) = \begin{bmatrix} F_{\text{up}}(t_n) & F_{\text{down}}(t_n) \end{bmatrix}^\mathsf{T} \) represents the actual hydrodynamic forces,  
and \( \boldsymbol{m}_p = \begin{bmatrix} m_{p,\text{up}} & m_{p,\text{down}} \end{bmatrix}^\mathsf{T} \) is the vector of physical masses of the upstream and downstream pipe models, respectively.
Lightweight materials such as carbon fiber are employed in the construction of the pipe models to effectively reduce their physical masses and thereby minimize the impact of inertial forces on the measurements.  

To eliminate the influence of inertial forces from the measured hydrodynamic loads, the input mass $m_{\text{in}}$ in the numerical system is set to \( m + m_{p,\text{up}} + m_{p,\text{down}} \). Combining with \cref{eq:force_total}, the following governing equation of motion is obtained:
\begin{equation}
    (m - m_{p,\text{up}}-m_{p,\text{down}}) \ddot{y}(t_n) + c \dot{y}(t_n) + ky(t_n) = F_{\text{up}}(t_n)+ F_{\text{down}}(t_n) -(m_{p,\text{up}} + m_{p,\text{down}}) \ddot{y}(t_n),
\label{eq:dynamic_equation1}
\end{equation}
where \( m \), \( c \) and \( k \) are the virtual total mass, damping and stiffness of the twin-pipe model, respectively. By adding the physical mass of each pipe model to the virtual total mass, the inertial force term in the equation can be eliminated, resulting in the equation of motion for the desired physical structure, as follows:
\begin{equation}
    m \ddot{y}(t_n) + c \dot{y}(t_n) + ky(t_n) = F_{\text{up}}(t_n)+ F_{\text{down}}(t_n).
\label{eq:dynamic_equation2}
\end{equation}

The compensation time delay method used in this study is the same as that of \cite{ren2024developing}, and the derivation process will not be repeated here. The compensation negative damping ratio $c_e$ in this study for the twin-pipe model is
\begin{equation}
c_e = (m_{p,\text{up}} + m_{p,\text{down}}) \omega^2 \Delta t
\end{equation}
where \( \Delta t \) is the time delay, which can be measured from the time difference between the command signal and the execution signal. This equivalent negative damping is directly added to the damping term in \cref{eq:dynamic_equation1}, leading to an effective input damping given by $c_{\text{in}}$ = $c - c_e$.

The above VFP method facilitates the simultaneous simulation of the motion response of a twin-pipe model, where both upstream and downstream pipes experience hydrodynamic loads, with the influence of inertial forces and time delay eliminated. It also enables high-frequency measurement of hydrodynamic forces on both pipes, supporting detailed analysis of their hydrodynamic characteristics and interaction mechanisms.

\section{Description of experimental setup and VPF validation}\label{setup_vpf}

\subsection{Experimental setup and test matrix}

\begin{figure}[htbp!]
	\centering
	\includegraphics[width=1\textwidth]{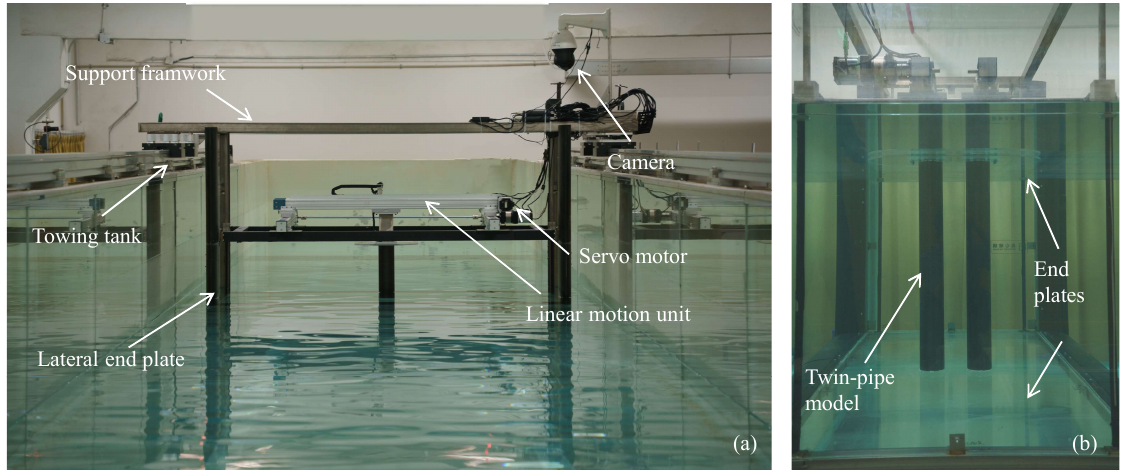}
	\caption{Experimental setup for FIV tests on a twin-pipe model. 
    (a) Overview of the setup in a towing tank. 
    (b) Side view of the twin-pipe model with end plates.}
	\label{setup}
\end{figure}

The experiments were carried out in a towing tank at the Institute of Marine Equipment, Shanghai Jiao Tong University, as shown in \cref{setup}. The dimensions of the towing tank are \SI{26}{\meter} $\times$ \SI{2.5}{\meter} $\times$ \SI{1.6}{\meter} (\text{length} $\times$ \text{width} $\times$ \text{depth}). The experimental setup was mounted on a support framework attached to the towing carriage, as shown in \cref{setup}(a). The twin-pipe model was towed to simulate uniform flow, allowing for a controlled investigation of its FIV behavior in cross-flow (CF) direction. \cref{setup}(b) presents a side view of the experimental setup, illustrating the twin-pipe model with end plates. The twin-pipe model consists of two identical carbon fiber pipes that are rigidly connected, each measuring \( L = \SI{0.8}{\meter} \) in length and \( D = \SI{0.1}{\meter} \) in diameter, giving an aspect ratio of 8. The mean roughness of the pipe is \( 8.82 \times 10^{-6} \). The center-to-center distance $G$ of the twin-pipe model is $2D$.

The mass ratio is defined as \( m^* = m / m_\Delta \), where \( m \) is the total mass of the twin-pipe model, \( m_\Delta =  \pi\rho L D^2/2 \) is the total displaced mass of the twin-pipe model, and \( \rho \) is the density of water.
Three representative values of 1.0, 2.4, and 10.0 were chosen. These mass ratios were selected based on their prevalence in single-pipe FIV studies, covering a low, intermediate, and high range to systematically analyze mass ratio effects on the twin-pipe model. 
Three structural damping ratios ($\zeta$) in air of 0.000, 0.001, and 0.005 were used to examine their influence on the FIV response. With the aid of the VPF, the physical mass of the twin-pipe model was no longer constrained by material properties and the structural damping ratio can also be precisely adjusted.
The reduced velocity was defined as \( U_R = U / (f_n D) \), where \( U \) is the uniform flow velocity and $f_n = (1 / 2\pi) \sqrt{k / (m+m_a)}$ is the natural frequency of the flexibly mounted twin-pipe model in water, and $m_a$ is the added mass of the twin-pipe model, defined as $m_a = C_m\pi\rho D^2L  /2$, $C_m$ is set to be 1. By changing the structural stiffness \( k \), \( U_R \) was modified while keeping the flow velocity \( U \) constant, resulting in a fixed Reynolds number of \( 2.0 \times 10^4 \). The \( U_R \) range was arranged to cover the full development of the FIV response, with a typical increment of 1. In regions with significant response variation, additional intermediate points were added. A summary of all test conditions is provided in \cref{tab:test_matrix}.

\begin{table}[htbp!]
\centering
\caption{Test matrix for FIV experiments of the twin-pipe model}
\begin{tabular*}{0.7\textwidth}{@{\extracolsep{\fill}}cccc}
\toprule
Mass ratio \( m^* \) & Damping ratio \( \zeta \) & Reduced velocity \( U_R \) & Reynolds number Re \\
\midrule
1.0  & 0.000   & \( 3{:}1{:}32 \) & \( 2.0 \times 10^4 \) \\
1.0  & 0.001   & \( 3{:}1{:}32 \) & \( 2.0 \times 10^4 \) \\
1.0  & 0.005   & \( 3{:}1{:}32 \) & \( 2.0 \times 10^4 \) \\
2.4  & 0.000   & \( 3{:}1{:}23 \) & \( 2.0 \times 10^4 \) \\
10.0 & 0.000   & \( 3{:}1{:}20 \) & \( 2.0 \times 10^4 \) \\
\bottomrule
\end{tabular*}
\label{tab:test_matrix}
\end{table}

\subsection{Validation of experimental setup and VPF performance}

\begin{figure}[htbp!]
	\centering
	\includegraphics[width=1\textwidth]{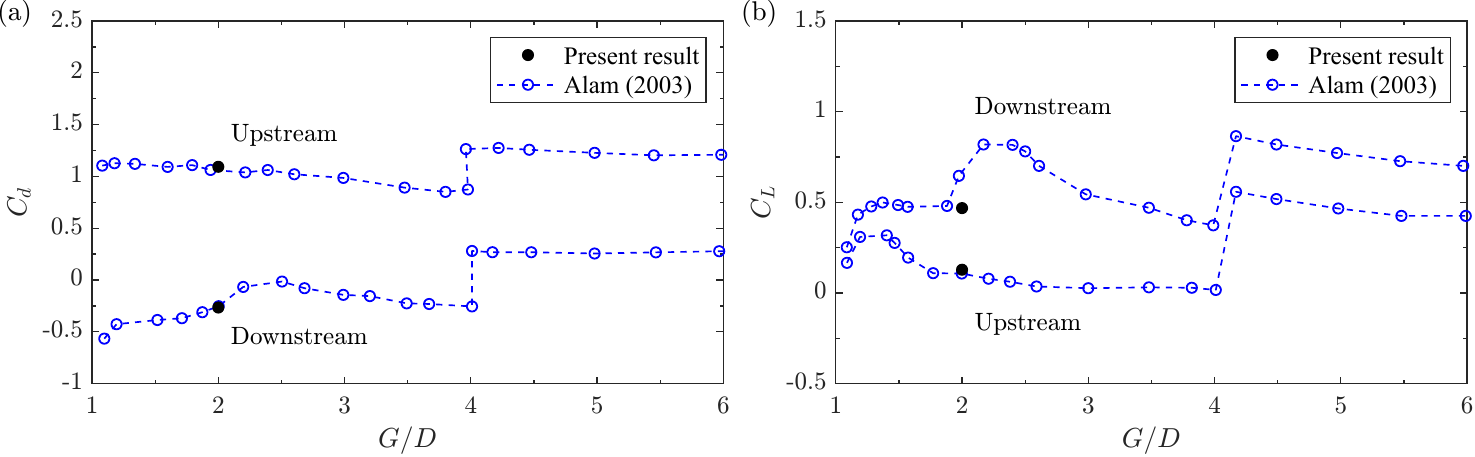}
	\caption{Mean drag force coefficient \( C_d \) and fluctuating lift force coefficient \( C_L \) for the twin-pipe model as a function of \( G/D \). (a) shows \( C_d \) for the upstream and downstream pipes, and (b) presents \( C_L \). The present results (black dots) are compared with the experimental data from \cite{alam2003fluctuating} (blue circles).
 }
	\label{Twin_station}
\end{figure}

To verify the reliability of the towing carriage and the force sensors, stationary towing tests were first conducted on the twin-pipe model.  \cref{Twin_station}(a) presents the mean drag coefficient \( C_d \) as a function of the spacing ratio \( G/D \), and \cref{Twin_station}(b) shows lift coefficient \( C_L \) versus \( G/D \). The \( C_d \) and \( C_L \) are defined in \cref{eq:cd_cl}:

\begin{equation}
C_d = \frac{\overline{F_d(t)}}{\frac{1}{2} \rho L U^2}, \quad
C_L = \frac{\sqrt{2} F_{L,\mathrm{rms}}}{\frac{1}{2} \rho L U^2},
\label{eq:cd_cl}
\end{equation}
where \( \overline{F_d(t)} \) is the time-averaged drag force, \( F_{L,\mathrm{rms}} \) is the root-mean-square value of the lift force, \( U \) is the velocity of the uniform flow, \( \rho \) is the density of water, \( L \) is the length of the pipe, and \( D \) is the diameter of the pipe.

The results from the present study exhibit good agreement with those reported by \cite{alam2003fluctuating}, demonstrating the reliability of the experimental setup. Since the primary objective of this study is to investigate the effects of mass and damping on the FIV response, a fixed spacing ratio of \( G/D = 2 \) was adopted. The Reynolds numbers for the present tests and those in \cite{alam2003fluctuating} are \( 2.0 \times 10^4 \) and \( 6.4 \times 10^4 \), respectively. It can be seen that this difference in Reynolds number has a relatively small effect on the hydrodynamic coefficients, ensuring the validity of the comparison.

\begin{figure}[htbp!]
	\centering
	\includegraphics[width=1\textwidth]{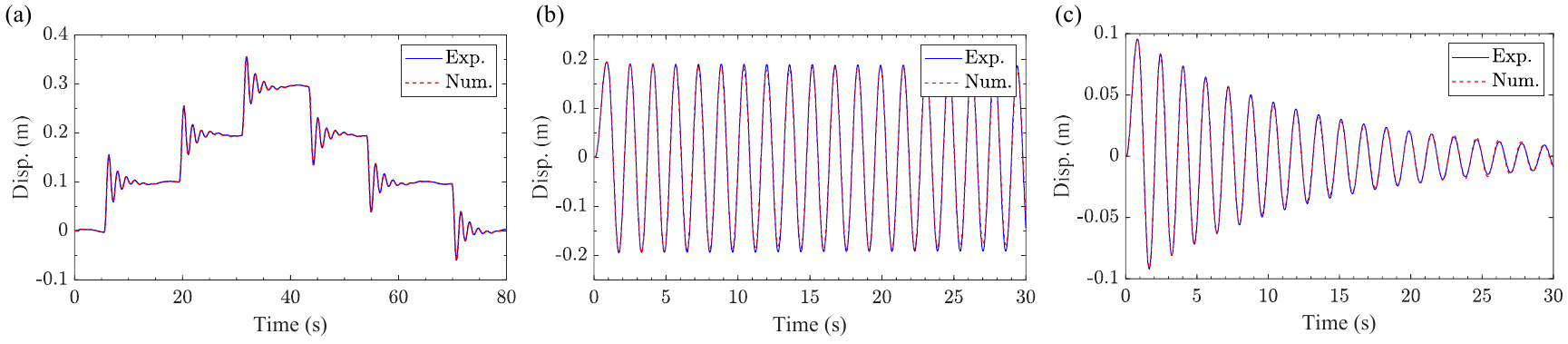}
	\caption{Validation of VPF on simulating mass, spring stiffness, and damping ratio for an elastic system in air. (a) Response under step loading and unloading with $m=\SI{6.28}{kg}$, $k=\SI{100}{N/m}$, $\zeta = 0.150$, (b) free decay response with $m=\SI{6.28}{kg}$, $k=\SI{100}{N/m}$, $\zeta = 0.000$ and (c) free decay response with $m=\SI{6.28}{kg}$, $k=\SI{100}{N/m}$, $\zeta = 0.020$. The blue solid lines represent experimental results, and the red dashed lines indicate the theoretical solutions. }
	\label{Airvalidation}
\end{figure}

Further validation tests were conducted to assess whether the VFS system can accurately replicate the dynamic response of a 1-DoF spring system, especially its ability to edit structural parameters. Static load tests, through step loading and unloading of heavy objects, were conducted to verify the stiffness control. \cref{Airvalidation} (a) illustrates the whole static load test case process with $m=\SI{6.28}{kg}$, $k=\SI{100}{N/m}$, $\zeta = 0.150$, and exhibits good agreement between experiment and numerical simulation. To further validate damping control, free decay tests were performed. In these tests, the structure was displaced and then released, allowing it to oscillate freely without external excitation. The time histories of two free decay cases with damping ratios of $\zeta=$ 0.000 and 0.020 with $m=\SI{6.28}{kg}$ and $k=\SI{100}{N/m}$ are illustrated in \cref{Airvalidation} (b) and (c). Consistency between the actual results from the VPF and the numerical simulation can be observed. The error between the measured damping ratio and the set damping ratio was $4\%$. The matching oscillation periods in \cref{Airvalidation}(b) and (c) further confirm the accurate reproduction of the system's natural period, which is given by \( T_n = 2\pi \sqrt{m / k} \). Since the stiffness control has already been verified in \cref{Airvalidation}(a), this also indicates that the mass parameter can be precisely edited. 
The aforementioned results demonstrated that the VPF system performs accurately in terms of damping ratio, stiffness, and mass control.

\begin{figure}[htbp!]
	\centering
	\includegraphics[width=1\textwidth]{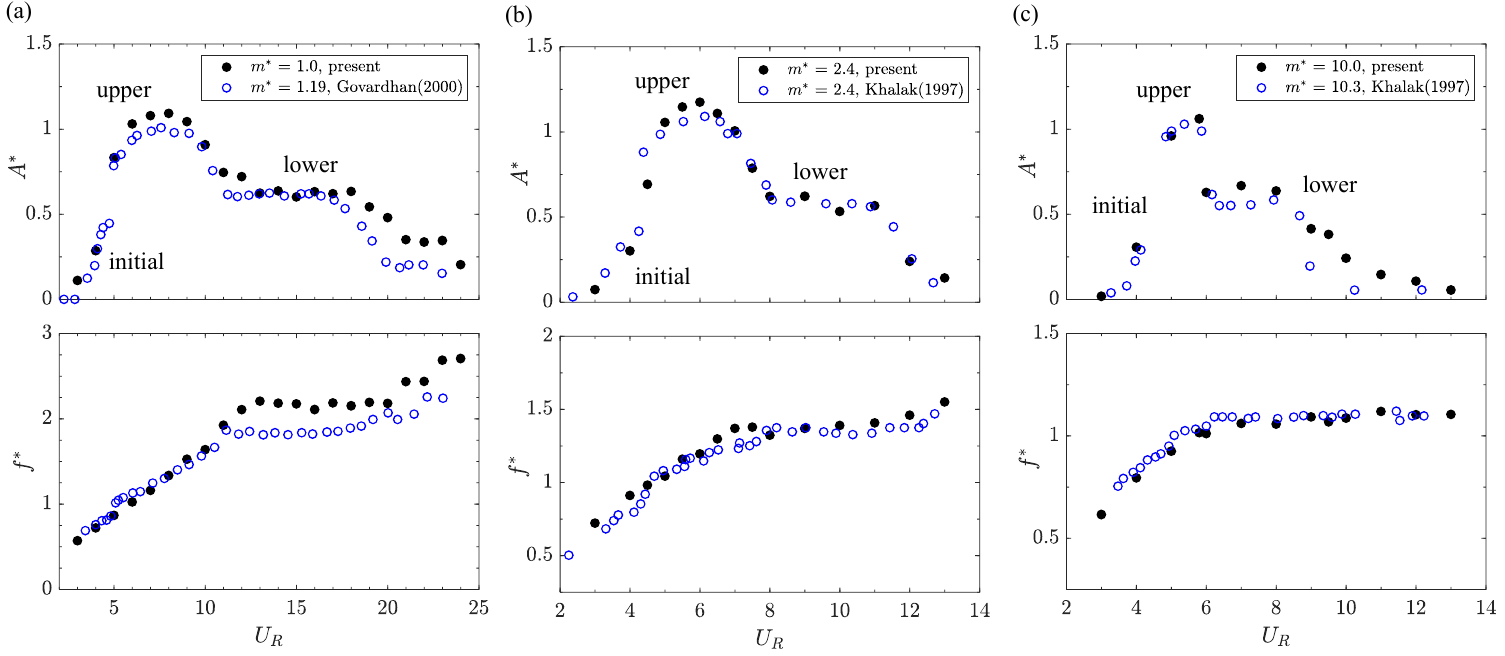}
	\caption{Comparison of \( A^* \) and \( f^* \) versus \( U_R \) for different mass ratios. (a) Present results for \( m^* = 1.0 \) compared with \cite{govardhan2000modes}, \( m^* = 1.19 \). (b) Present results for \( m^* = 2.4 \) compared with \cite{khalak1997investigation}, \( m^* = 2.4 \). (c) Present results for \( m^* = 10.0 \) compared with \cite{khalak1997investigation}, \( m^* = 10.3 \). }
	\label{Wetvalidation}
\end{figure}

Building upon the successful validation in air, the VPF system was further verified in water conditions to ensure its reliability for formal FIV experiments. Single-pipe tests were performed and compared with previously published benchmark data. \cref{Wetvalidation} presents a comparison of the non-dimensional FIV amplitude ($A^* = \sqrt{2}y_{rms}/D$,  where $y_{rms}$ represents the root-mean-square (rms) value of the displacement) and the non-dimensional dominant vibration frequency ($f^* = f_d/f_n$, where \( f_d \) is the dominant FIV frequency, and \( f_n \) is the natural frequency of the system in water) as functions of \( U_R \) for different mass ratios.  \cref{Wetvalidation}(a) compares the case of \( m^* = 1.0 \) with \cite{govardhan2000modes} (\( m^* = 1.19 \)), \cref{Wetvalidation}(b) presents \( m^* = 2.4 \) against \cite{khalak1997investigation} (\( m^* = 2.4 \)), and \cref{Wetvalidation}(c) examines \( m^* = 10.0 \) in comparison with \cite{khalak1997investigation} (\( m^* = 10.3 \)).  The Reynolds number in this study is kept constant at \( Re = 2.0 \times 10^4 \). Although the Reynolds numbers were not explicitly provided in the reference literature, they were estimated to lie within the range of \( 10^3\) to \(10^4 \) based on the reported experimental parameters.

The upper subfigures in \cref{Wetvalidation} show the variation of the non-dimensional amplitude \( A^* \) with reduced velocity \( U_R \), clearly capturing the initial branch, upper branch, and lower branch as defined in the work of \cite{williamson2008brief}. The lower subfigures present the corresponding non-dimensional dominant frequency ratio \( f^* \), revealing the characteristic lock-in behavior. A detailed comparison with previously published results \citep{govardhan2000modes, khalak1997investigation} confirms the strong agreement of the present experimental results. The present results not only follow the same trends but also match well in response values for both \( A^* \) and \( f^* \) across all mass ratios. For \( m^* = 1.0 \), the amplitude response shows a slight shift toward higher reduced velocities in the lower branch, and the lock-in frequency \( f^* \) is slightly higher compared to the results of \cite{govardhan2000modes} for \( m^* = 1.19 \). These differences are likely attributed to the slightly lower mass ratio in the present study. Despite such variations, the observed response trends remain consistent with established FIV behavior \citep{govardhan2006defining}, thereby confirming the reliability of the VPF-based setup for subsequent experimental investigations.

\section{Results and discussion}\label{results}
\subsection{Mass effects on FIV amplitude and frequency}

Systematic FIV tests were conducted on a twin-pipe model with mass ratios of 1.0, 2.4, and 10.0 under uniform flow conditions. For each mass ratio, the towing velocity was maintained at 0.2 m/s, corresponding to a Reynolds number of \( 2.0 \times 10^4 \), with a damping ratio set to zero. The tested \( U_R \) covered the entire process of FIV amplitude development, as shown in \cref{tab:test_matrix}, by varying the natural frequency \( f_n \) through stiffness adjustments within VPF.

\begin{figure}[htbp!]
	\centering
	\includegraphics[width=1\textwidth]{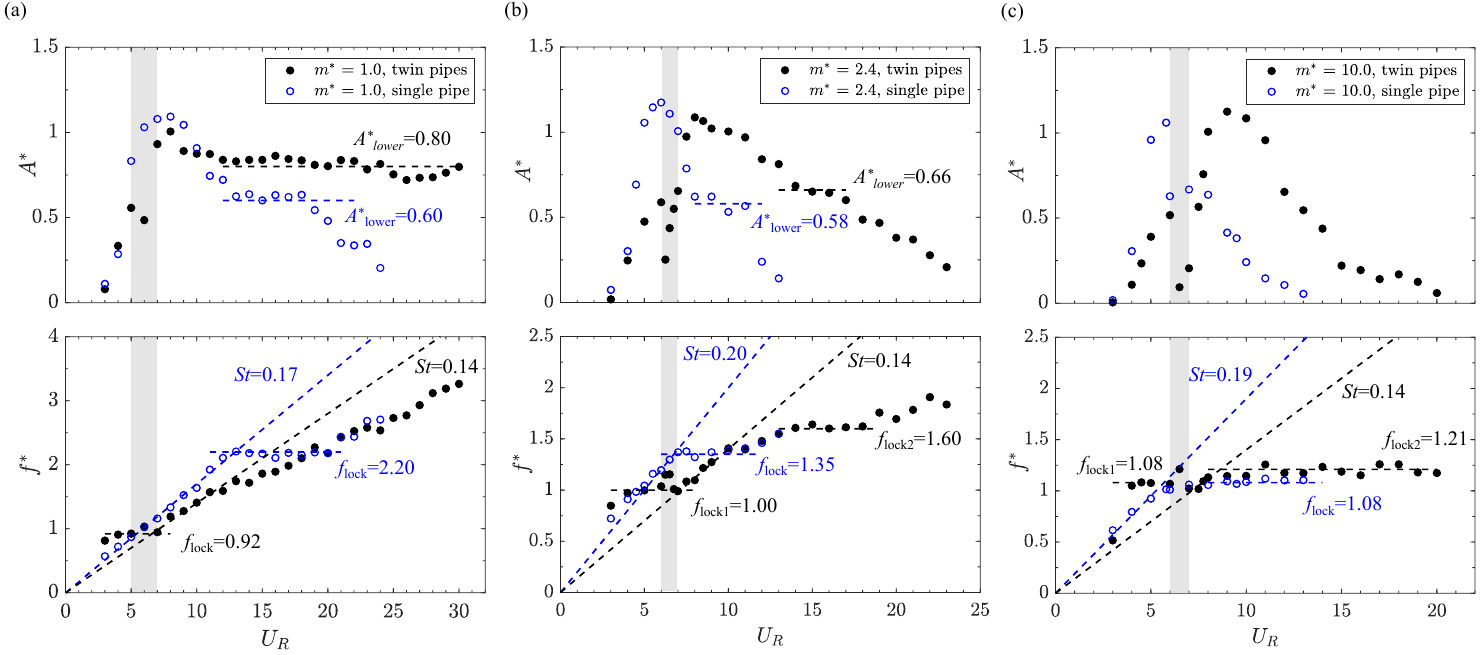}
	\caption{Comparison of FIV responses between twin-pipe model (black dots) and single-pipe model (blue open circles) for different mass ratios. (a) \( m^* = 1.0 \), (b) \( m^* = 2.4 \), and (c) \( m^* = 10.0 \). The upper row presents \( A^* \) versus  \( U_R \), and the lower row shows the \( f^* \) versus \( U_R \). The shaded region indicates the amplitude drop region.}
	\label{massAf}
\end{figure}

\cref{massAf} presents the non-dimensional FIV amplitude \(A^*\) and dominant frequency ratio \(f^*\) as functions of the reduced velocity \(U_R\) for both single-pipe and twin-pipe systems across three different mass ratios.
For the single-pipe model, the typical FIV response depends strongly on the mass ratio. At low mass ratios, the FIV development typically consists of three distinct stages: the initial branch, where the amplitude grows gradually toward its peak, the upper branch, corresponding to the region of large and sustained amplitude, and the lower branch, where the amplitude decreases after the peak and gradually vanishes \citep{williamson2008brief}. As the mass ratio increases, the response is dominated by the initial and lower branches.
The twin-pipe model exhibits similar behavior, but with notable differences. Most prominently, the synchronization region in the twin-pipe system extends over a broader \(U_R\) range.
The twin-pipe model consistently shows higher amplitudes than the single-pipe model in the lower branch. At \(m^* = 1.0\), the stable lower branch amplitude reaches approximately \(0.80D\) for the twin-pipe model, compared to \(0.60D\) for the single-pipe model. Similarly, for \(m^* = 2.4\), the twin-pipe model shows \(0.66D\), whereas the single-pipe model remains at \(0.58D\). For \(m^* = 10.0\), no stable amplitude is observed in the lower branch for either case.
Additionally, for all mass ratios, a distinct phenomenon appears in the twin-pipe response near \(U_R = 6\), where the amplitude exhibits a sudden local drop. This feature, defined as the ``amplitude drop'' is highlighted in the shaded region of \cref{massAf}. A similar observation was also reported by \cite{shen2024experimental}, where a comparable amplitude drop occurred in the FIV response of rigidly coupled tandem twin pipes with a spacing ratio of $2D$. Such a drop is absent in the single-pipe case.

In terms of frequency characteristics, a notable distinction emerges in the amplitude drop region. The twin-pipe model exhibits a frequency lock-in region where the non-dimensional frequency stabilizes around 1.0, whereas no such lock-in behavior is observed in the single-pipe model. Regarding the Strouhal number (St), its value for the twin-pipe model remains relatively consistent at approximately 0.14. In contrast, the single-pipe model exhibits slightly higher St values across different mass ratios than those of the twin-pipe model. At higher \(U_R\), the lock-in frequency in the twin-pipe model is slightly higher than that of the single-pipe model.

\cref{massAf} further reveals the mass ratio effects on FIV responses of the twin-pipe model. For \( m^* = 1.0 \), the response amplitude remains nearly constant for \( U_R > 10 \) and does not exhibit a tendency to decrease, even at \( U_R = 30 \). This sustained high amplitude may be attributed to the ``resonance forever'' phenomenon observed in single-pipe models by \cite{govardhan2002resonance}, suggesting that a similar mechanism may exist in the twin-pipe model. For a single-pipe model, when \( m^* \) is below a critical mass ratio of 0.54, the vibration amplitude does not decay with increasing reduced velocity. Given that \( m^* = 1.0 \) in the present twin-pipe model, it is likely that this value falls below the critical mass ratio for twin-pipe models, leading to the observed persistent vibration amplitude.
However, as the mass ratio increases, this behavior changes, and the response amplitude progressively decreases with increasing \( U_R \). The reduction in amplitude becomes more pronounced at higher mass ratios. For instance, at \( m^* = 2.4 \), the amplitude decreases to approximately \( 0.1D \) at \( U_R = 24 \), whereas at \( m^* = 10.0 \), the amplitude becomes negligible by \( U_R = 15 \). Additionally, in the amplitude drop region, the extent of amplitude reduction increases with mass ratio. At \( m^* = 1.0 \), the amplitude only decreases to about \( 0.5D \), while at \( m^* = 10.0 \), it drops to nearly \( 0.1D \). The effect of mass ratio on dominant vibration frequency is minor at low \( U_R \), where the dominant frequency follows the \( St = 0.14 \) trend. However, at larger \( U_R \), mass ratio has a stronger influence. At \( m^* = 1.0 \), there is almost no well-defined lock-in region, and the frequency gradually increases with \( U_R \). At \( m^* = 2.4 \), the non-dimensional dominant frequency locks near 1.60 within \( 12 < U_R < 18 \). At \( m^* = 10.0 \), the lock-in range extends to \( 10 < U_R < 20 \), with a lower lock-in frequency of approximately 1.21.

\subsection{Mass effects on FIV hydrodynamics}

To investigate the hydrodynamic features of the twin-pipe model under FIV, the mean drag coefficient \(C_d\), excitation coefficient \(C_v\), and added mass coefficient \(C_m\) for each pipe were extracted and analyzed individually. The \( C_d \) of each pipe is calculated by \cref{eq:cd_cl}.
To extract the excitation and added mass coefficients, the lift force \( F_L \) is further decomposed into two components: one in phase with velocity, referred to as the excitation force, and the other in phase with acceleration, referred to as the added mass force \citep{song2016investigation}. The corresponding formulation is given as:

\begin{equation}
    F_L(t) = \frac{1}{2\sqrt{2} \, \dot{y}_{\text{rms}}} \rho D L U^2 C_v \dot{y}(t) - \frac{\pi}{4} D^2 L \rho C_m \ddot{y}(t),
\end{equation}
The \( C_v \) and \( C_m \) are obtained using the least squares method based on the measured lift force.

\begin{figure}[htbp!]
	\centering
	\includegraphics[width=1\textwidth]{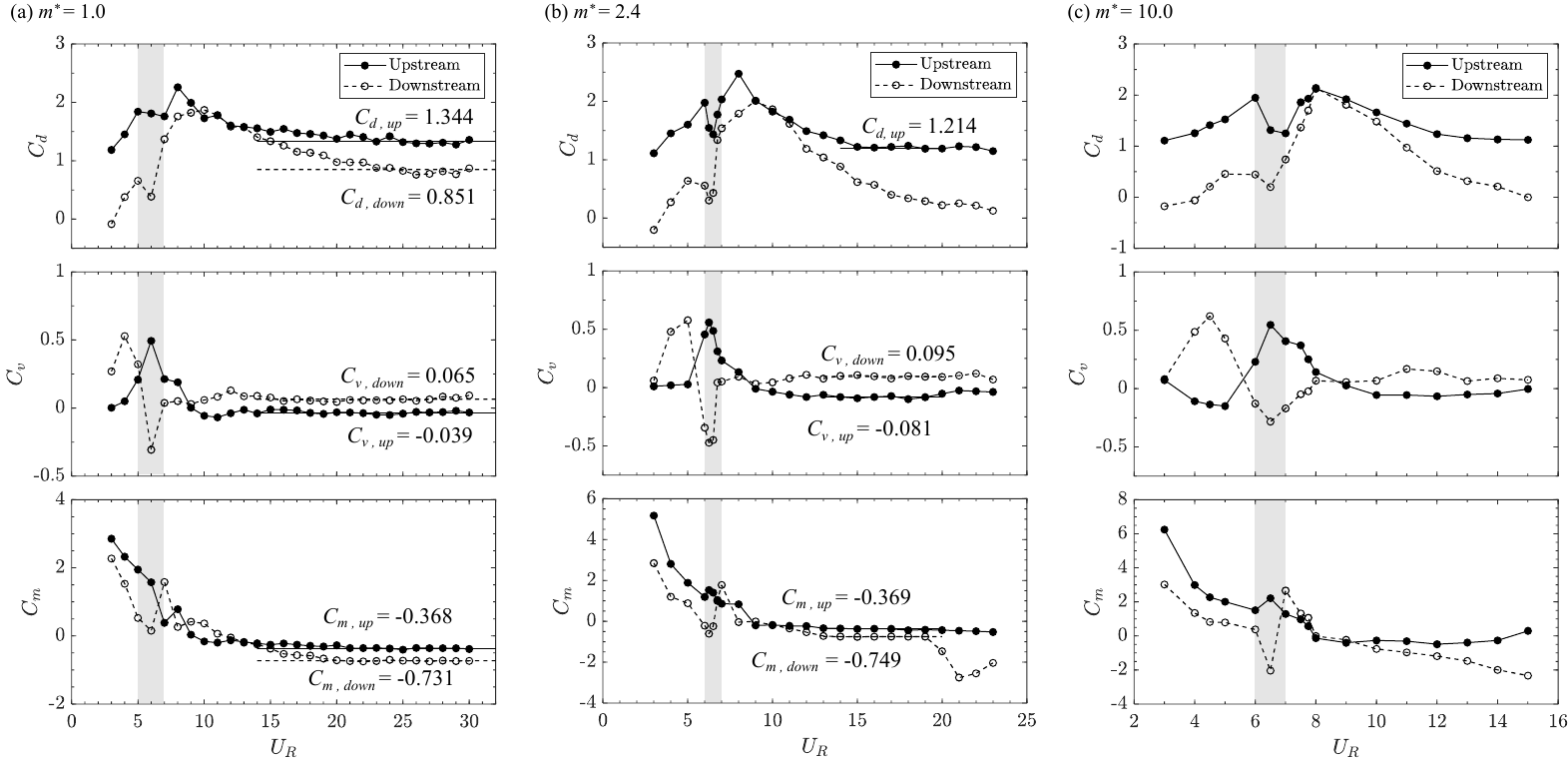}
	\caption{Variation of \( C_d \), \( C_v \), and \( C_m \) with \( U_R \) for both upstream and dwonstream pipes at different mass ratios. (a) \( m^* = 1.0 \), (b) \( m^* = 2.4 \), and (c) \( m^* = 10.0 \). The solid and open markers represent the upstream and downstream pipes, respectively. The shaded region highlights the amplitude drop region.}
	\label{massCDCL}
\end{figure}

\cref{massCDCL} shows the variations of \(C_d\), \(C_v\), and \(C_m\) with \(U_R\) for both the upstream and downstream pipes at three different mass ratios: (a) \(m^* = 1.0\), (b) \(m^* = 2.4\), and (c) \(m^* = 10.0\). For the case of \(m^* = 10.0\), the vibration amplitude becomes extremely weak at high reduced velocities, resulting in multi-frequency behavior in the force signals. As a result, the extracted hydrodynamic coefficients in this range are no longer reliable and are therefore not shown in the figure.

The first row of \cref{massCDCL} presents the results of \(C_d\) of each pipe.
A clear distinction emerges between upstream and downstream pipes. The upstream pipe generally experiences higher \(C_d\), as it is directly exposed to the incoming flow, whereas the downstream pipe is partially sheltered. Across different mass ratios, the overall variation trends of \( C_d \) with \( U_R \) are largely similar outside the stable region. The most distinct differences emerge in the region where drag coefficients reach steady values. At \( m^* = 1.0 \), both the upstream and downstream pipes exhibit relatively high and stable \( C_d \) values over an extended range of \( U_R \), while higher mass ratios (\( m^* = 2.4 \) and \( m^* = 10.0 \)) lead to lower stable values and narrower stability ranges.
It has been proved that FIV can significantly amplify the mean drag coefficient \( C_d \) and this amplification is positively correlated with the vibration amplitude \citep{zhao2023drag, deng2020experimentalD, vandiver1983drag}. The larger and more persistent vibrations at lower mass ratios enhance the \(C_d\). In contrast, higher mass ratios result in earlier FIV decay, leading to reduced and less sustained \( C_d \).
Within the amplitude drop region, local reductions in \( C_d \) are observed for all mass ratios, aligning with the sudden drop in vibration amplitude.

The excitation coefficient \( C_v \), representing the component of lift in phase with structural velocity, provides insight into the energy transfer during FIV. As shown in the second row of \cref{massCDCL}, all mass ratios exhibit a sharp transition in \( C_v \) within the amplitude drop region. In this region, the upstream and downstream pipes display opposite signs of \( C_v \), indicating strong phase desynchronization and energy imbalance between the two pipes. The pronounced negative excitation in the downstream pipe leads to energy dissipation within the system, resulting in a noticeable drop in vibration amplitude.
At higher reduced velocities, excitation coefficients of both pipes gradually approach zero. For \( m^* = 1.0 \) and \( m^* = 2.4 \), a relatively steady \( C_v \) stage is observed, while for \( m^* = 10.0 \), the force signal becomes unstable due to weak vibrations.

The added mass coefficient \( C_m \), representing the component of the lift force in phase with structural acceleration, is shown in the third row of \cref{massCDCL}. At lower reduced velocities, all cases exhibit a rapid decreasing trend in \( C_m \), to more stabilized hydrodynamic behavior at higher \( U_R \). Within the amplitude drop region, deviations from the overall decreasing trend are observed in \( C_m \).
At higher \( U_R \), both the upstream and downstream pipes tend to reach a stable state, particularly for lower mass ratios of \( m^* = 1.0 \) and \( m^* = 2.4 \).  Notably, the sum of the stabilized added mass coefficients \( C_m \) for the two pipes remains nearly constant across different mass ratios, with values approximately \(-1.10\) for \( m^* = 1.0 \) and \(-1.12\) for \( m^* = 2.4 \).  
In all cases where a stable \( C_m \) region is observed, the total added mass coefficient consistently converges to around \(-1.1\).  
This observation aligns with the critical mass ratio concept proposed in previous single-pipe studies \citep{govardhan2002resonance}, which suggests that when the sum of structural mass and added mass approaches zero, the system enters a state of continuous resonance known as resonance forever. In this condition, the critical mass ratio is given by $m^*_{crital}=-c_m$, and the vibration does not decay with increasing \( U_R \).
In the present twin-pipe model, the total \( C_m \) stabilizes around 1.1, indicating a critical mass ratio of approximately 1.1. The case of \( m^* = 1.0 \), which is below the critical mass ratio, shows the resonance forever phenomenon, characterized by sustained oscillations with increasing \( U_R \). This confirms that the critical mass effect observed in single-pipe systems also applies to rigidly coupled twin-pipe systems.

Unlike a single pipe, the rigidly coupled twin-pipe model exhibits a unique hydrodynamic behavior arising from the interaction between the upstream and downstream pipes. Specifically, differences in lift forces between the two pipes can generate torsional moments on the structure, while differences in drag forces can lead to tensile or compressive interaction along the in-line (IL) direction. In addition, the total drag acting on the coupled system also varies with \( U_R \), reflecting the total in-line resistance on the structure. The total drag coefficient
(\(C_{d,\text{total}} \)), differential drag coefficient (\( C_{d,\Delta} \))), and torsional coefficient (\( C_T \)) are defined as follows:

\begin{align}
    C_{d,\text{total}} &= \frac{\overline{F_{d,\text{up}}(t) + F_{d,\text{down}}(t)}}{\frac{1}{2} \rho D L U^2 }, \label{eq:Cdtotal} \\
    C_{d,\Delta}  &= \frac{\overline{F_{d,\text{up}}(t) - F_{d,\text{down}}(t)}}{\frac{1}{2} \rho D L U^2 }, \label{eq:CR} \\
    C_T &= \frac{T}{\frac{1}{4} \rho D L G U^2} = \frac{2\sqrt{2} \, (F_{L,\text{up}}(t) - F_{L,\text{down}}(t))_{\text{rms}}}{\rho D L U^2}, \label{eq:CT}
\end{align}
where \( F_{d,\text{up}}(t) \) and \( F_{d,\text{down}}(t) \) are the instantaneous drag forces, and \( F_{L,\text{up}}(t) \) and \( F_{L,\text{down}}(t) \) are the instantaneous lift forces, acting on the upstream and downstream pipes, respectively. \( T \) is torsional moment, calculated as \( T = \sqrt{2} \cdot (F_{L,\text{up}}(t) - F_{L,\text{down}}(t))_{\text{rms}} \cdot G/2 \). \( \overline{(\cdot)} \) denotes time-averaging, and \( (\cdot)_{\text{rms}} \) represents the root-mean-square value.

\begin{figure}[htbp!]
	\centering
	\includegraphics[width=1\textwidth]{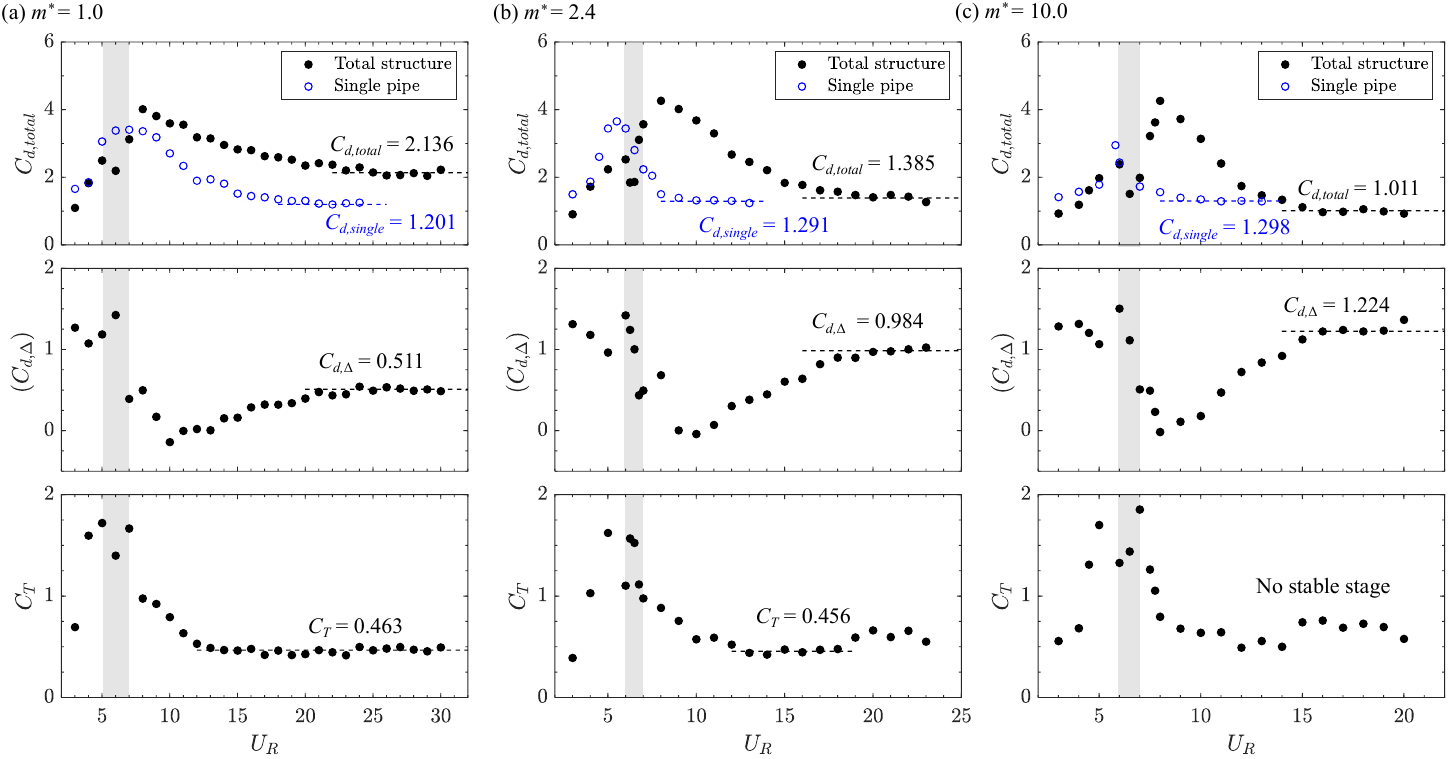}
	\caption{Variation of \( C_{d,\text{total}} \), \( C_{d,\Delta} \), and \( C_T \) with \( U_R \) for the twin-pipe model at different mass ratios. (a) \( m^* = 1.0 \), (b) \( m^* = 2.4 \), and (c) \( m^* = 10.0 \). The black solid circles represent the total structure, and the blue open circles denote the single-pipe case for comparison. }
	\label{massCT}
\end{figure}

\cref{massCT} presents the variations of the \( C_{d,\text{total}} \), \( C_{d,\Delta} \), and \( C_T \) with \( U_R \) at three mass ratios (\( m^* = 1.0, 2.4\) and 10.0). The first row of \cref{massCT} shows the variation of \( C_{d,\text{total}} \) with \( U_R \), for reference, the corresponding single-pipe results are also included.
In general, the twin-pipe model exhibits a higher \( C_{d,\text{total}} \) than the single-pipe model, especially at the peak drag region. Notably, in the amplitude drop region, all twin-pipe cases exhibit a sudden reduction in drag, which is not observed in the single-pipe model. At higher reduced velocities, \( C_{d,\text{total}} \) gradually stabilizes for all mass ratios. At \( m^* = 1.0 \), the final stabilized value is 2.136 for the twin-pipe model versus 1.201 for the single pipe, representing a 77.8\% increase. As the mass ratio increases, this gap narrows. At \( m^* = 2.4 \), the final \( C_{d,\text{total}} \) is 1.385 for the twin pipe and 1.291 for the single pipe, representing a 7.3\% increase. When \( m^* = 10.0 \), the stabilized drag coefficient for the twin pipe drops to 1.011, which is even lower than the single-pipe value of 1.298.

The second row of \cref{massCT} presents the variation of \( C_{d,\Delta} \) with \( U_R \). According to \cref{eq:CR}, a positive \( C_{d,\Delta} \) indicates a compressive effect along the in-line direction, while a negative value corresponds to a tensile effect. In general, \( C_{d,\Delta} \) remains positive across most of the range, indicating that the twin-pipe model primarily experiences compressive hydrodynamic interaction under FIV. This compressive effect first weakens and then strengthens as \( U_R \) increases, eventually stabilizing at a nearly constant value. In the amplitude drop region, \( C_{d,\Delta} \) increases sharply and reaches a local maximum, with peak values of 1.61, 1.43, and 1.50 for \( m^* = 1.0 \), 2.4, and 10.0, respectively. These peak values are similar, suggesting limited sensitivity of the maximum compressive force to the mass ratio.
Around \( U_R = 10 \), \( C_{d,\Delta} \) reaches its minimum and slightly negative value, coinciding with the maximum FIV amplitude response. This suggests that the compressive effect disappears at this point and a weak tensile force may even occur.
The influence of mass ratio is most evident in the stabilized regime. The stable \( C_{d,\Delta} \) increase with mass ratio, from 0.511 at \( m^* = 1.0 \) to 0.984 at \( m^* = 2.4 \), and further to 1.224 at \( m^* = 10.0 \), indicating stronger compressive interactions at higher mass ratio.

The third row of \cref{massCT} shows the variation of the torsional coefficient \( C_T \). \( C_T \) exhibits a rising-then-falling trend with increasing \( U_R \), with the maximum value occurring around \( U_R = 6 \). The peak values are approximately 1.80 across all three mass ratios, indicating limited mass ratio dependence in the maximum \( C_T \).
Within the amplitude drop region, the trend of \( C_T \) experiences a noticeable shift. For \( m^* = 1.0 \) and \( m^* = 10.0 \), \( C_T \) slightly decreases, whereas for \( m^* = 2.4 \), it slightly increases. As previously discussed, this region corresponds to the condition where the excitation coefficient \( C_v \) for the upstream and downstream pipes shows opposite signs, indicating a sudden change in phase difference. This abrupt shift can lead to a jump in the lift force difference in \cref{eq:CT}. Such phase shifts can increase or decrease \( C_T \), but in either case, they change the prior trend.
The influence of mass ratio is primarily reflected at higher \( U_R \). For \( m^* = 1.0 \), a stable stage appears for \( U_R > 12 \), with \( C_T \) stabilizing at 0.463. This stage coincides with the stable amplitude regime of the FIV response. For \( m^* = 2.4 \), a similar stable region exists between \( U_R = 12 \) to 18, with a steady \( C_T \) value of 0.456, which is close to that of the \( m^* = 1.0 \) case. This also corresponds to the stable amplitude range of FIV. In contrast, the \( m^* = 10.0 \) case shows no stable stage in \( C_T \), consistent with the absence of a stable FIV amplitude regime. 
In summary, the mass ratio influences the existence and extent of the stable region for \( C_T \), but appears to have limited effect on the final stable value itself.

In summary, the mass ratio has a significant impact on the FIV responses and hydrodynamic interactions of the twin-pipe model. These findings provide practical guidance for the engineering design of such configurations. Specifically, low mass ratios should be avoided to prevent the resonance forever phenomenon, while high mass ratios may induce strong compressive effects, requiring consideration in structural strength design.

\subsection{Damping effects on FIV responses and hydrodynamics}

\begin{figure}[htbp!]
	\centering
	\includegraphics[width=1\textwidth]{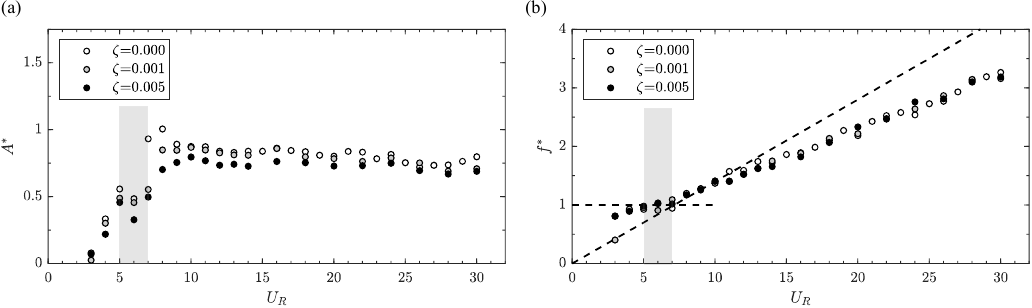}
	\caption{Comparison of FIV responses of the twin-pipe model at $m^*=1.0$ with different damping ratios. (a) Variation of \( A^* \) with \( U_R \) and (b) variation of \( f^* \) with \( U_R \). The open, gray, and black circles represent damping ratios of \(\zeta = 0.000, 0.001,\) and \( 0.005\), respectively. }
	\label{damping_effect_Af}
\end{figure}

\cref{damping_effect_Af} presents the variation of \( A^* \) and \( f^* \) with \( U_R \) for the twin-pipe model with different damping ratios (\( \zeta = 0.000, 0.001, 0.005 \)). To isolate the effect of damping, all results shown correspond to the case with the same mass ratio of \( m^* = 1.0 \).
As shown in \cref{damping_effect_Af}(a), increasing damping suppresses the vibration amplitude across the entire \( U_R \) range. Notably, it makes the amplitude drop more noticeable. At the minimum point within the drop region, \( A^* \) reduces from 0.485 at \(\zeta = 0.000\) to 0.445 at \(\zeta = 0.001\), a reduction of 8.2\%, and further to 0.327 at \(\zeta = 0.005\), representing a 32.6\% decrease. At \( U_R = 7 \), the zero-damping case has largely exited the drop with an amplitude of 0.931, while the lightly and heavily damped cases remain suppressed at 0.553 and 0.497, respectively.
Damping also significantly lowers the maximum vibration amplitude, which drops from 1.005 at zero damping to 0.849 and 0.795 as \(\zeta\) increases to 0.001 and 0.005, corresponding to reductions of 15.5\% and 20.9\%. In the high \( U_R \) regime (\( U_R > 10 \)), the amplitude in the  zero-damping case remains consistently 0.08 to 0.10\( D \) higher than that of the heavily damped case, indicating a sustained suppression effect.
As shown in \cref{damping_effect_Af}(b), the dominant frequency \( f^* \) remains similar across the damping ratios examined in this study. However, this conclusion is specific to the damping and mass ratio ranges considered, and further studies are needed to assess whether different behaviors emerge with higher damping or different mass ratios.
In general, increasing damping reduces the vibration amplitude and makes the amplitude drop more noticeable, while having little effect on the dominant frequency.

\begin{figure}[htbp!]
	\centering
	\includegraphics[width=1\textwidth]{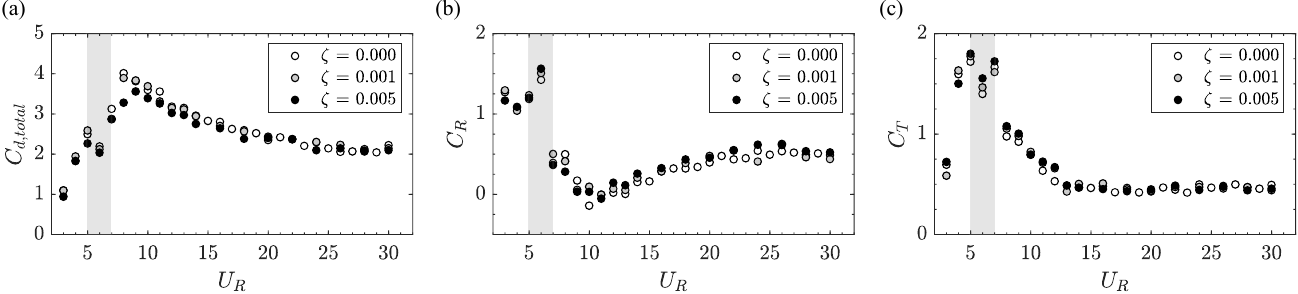}
	\caption{Variation of hydrodynamic coefficients of the twin-pipe model at $m^*=1.0$ with different damping ratios. (a) Variation of \( C_{d,\text{total}} \) with \( U_R \). (b) Variation of \( C_{d,\Delta} \) with \( U_R \). (c) Variation of \( C_T \) with \( U_R \). The open, gray, and black circles represent damping ratios of \(\zeta = 0.000, 0.001,\) and \( 0.005\), respectively.}
	\label{damping_effect_CT}
\end{figure}

\cref{damping_effect_CT} presents the variation of \( C_{d,\text{total}} \), \( C_{d,\Delta} \), and \( C_T \) with \( U_R \) for the twin-pipe model with different damping ratios (\( \zeta = 0.000, 0.001, 0.005 \)).
As shown in \cref{damping_effect_CT}(a), increasing damping reduces the overall drag force, especially near the amplitude drop region. Beyond this region, the effect of damping becomes less pronounced.
\cref{damping_effect_CT}(b) shows that the drag differential coefficient \( C_{d,\Delta} \) is only slightly affected by damping. Within the amplitude drop region, higher damping slightly enhances the local minimum of \( C_{d,\Delta} \). At larger \( U_R \), the \( C_{d,\Delta} \) tends to increase with damping, suggesting enhanced compressive interaction.
As shown in \cref{damping_effect_CT}(c), damping affects the torsional moment coefficient \( C_T \) at low reduced velocities. However, at higher \( U_R \), the influence of damping diminishes, and the curves for different damping ratios show very similar trends. Combined with the earlier analysis of mass ratio effects, this suggests that the stabilized value of \( C_T \) is largely independent of both mass ratio and damping ratio.
The results indicate that the effect of damping on hydrodynamic force coefficients is generally limited. A reduction is observed only in the total drag coefficient \( C_{d,\text{total}} \), while its influence on \( C_{d,\Delta} \) and \( C_T \) remains relatively minor.

In summary, the damping ratio primarily affects the vibration amplitude, which in turn influences the drag coefficient through the FIV amplification effect. In contrast, its impact on the dominant frequency, as well as on the differential drag coefficient and torsional coefficient, is relatively limited across the cases studied.

\section*{Conclusion}

This study develops a virtual physical framework (VPF) for the investigation of FIV on a rigidly coupled tandem twin-pipe model, and verifies its reliability by a serie of tests. The method allows flexible modification of structural mass, stiffness, and damping. Based on the extended VPF, systematic experiments were conducted to investigate the FIV responses and hydrodynamic characteristics of a twin-pipe model with varying mass ratios and damping ratios. For comparison, corresponding single-pipe FIV results were also included. The main findings are summarized as follows:

1) The twin-pipe model exhibits a broader synchronization region compared to the single pipe. As the mass ratio increases, the FIV amplitude easily weakens at lower $U_R$, and the lock-in frequency becomes lower. At \( m^* = 1.0 \), a “resonance forever” behavior is observed, with no distinct lock-in stage. Additionally, an ``amplitude drop'' near \( U_R = 6 \) is identified, attributed to strong phase opposition and energy dissipation associated with the downstream pipe.

2) The shielding effect from the upstream pipe results in a consistently lower mean drag coefficient of the downstream pipe. At higher reduced velocities, the wake effect becomes more prominent, leading to larger excitation coefficients of the downstream pipe. The influence of mass ratio on hydrodynamic coefficients also becomes more evident at high \( U_R \). For example, at \( m^* = 10 \), the drag, excitation, and added mass coefficients do not reach stable values. In contrast, for smaller mass ratios such as \( m^* = 1.0 \), these coefficients remain relatively steady. Notably, the total added mass coefficient for the two pipes stabilizes around –1.1, corresponding to a critical mass ratio of approximately 1.1.

3) The in-line hydrodynamic interaction between the two pipes is initially compressive, weakens with increasing \( U_R \), and reverses to slight repulsion near the peak FIV amplitude. With further increase in \( U_R \), the compressive effect strengthens again and eventually stabilizes, with stronger compressive effects and larger differential drag coefficients observed at higher mass ratios. The torsional moment coefficient increases with \( U_R \), then decreases and stabilizes around 0.46 at low mass ratios, while this stabilization is absent at high mass ratios. Notably, both the differential drag and torsional moment coefficients exhibit abnormal variations in the amplitude drop region.

4) The damping ratio mainly influences the vibration amplitude, particularly the maximum amplitude and the extent of the amplitude drop. Higher damping reduces overall response amplitudes, while its influence on dominant frequency is minimal. Among the hydrodynamic coefficients, only the drag coefficient shows limited sensitivity to damping, with a slight decrease as damping increases.

In summary, these findings provide practical guidance for the engineering design of twin-pipe structures. Specifically, low mass ratios should be avoided to prevent the resonance forever phenomenon that pose significant fatigue risks. In contrast, high mass ratios may induce strong compressive interactions between pipes, which must be considered in structural strength design, particularly the reliability of connectors.

The established hybrid experimental platform can be further used to investigate the sensitivity of structural responses to parameter variations under different spacing ratios, providing a more complete foundation for future engineering applications.

\section*{Acknowledgments}
The authors gratefully acknowledge the financial support from National Key Research and Development Program of China under Grant Number of 2022YFB2602800,
National Science Fund for Distinguished Young Scholars under Grant Number of 52425102,  National Natural Science Foundation of China under Grant Number of 52088102, the Xplorer Prize under Grant Number of XPLORER-2022-1037, and the Fundamental Research Funds for the Central Universities.

\newpage

\begin{thebibliography}{26}
\expandafter\ifx\csname natexlab\endcsname\relax\def\natexlab#1{#1}\fi
\providecommand{\url}[1]{\texttt{#1}}
\providecommand{\href}[2]{#2}
\providecommand{\path}[1]{#1}
\providecommand{\DOIprefix}{doi:}
\providecommand{\ArXivprefix}{arXiv:}
\providecommand{\URLprefix}{URL: }
\providecommand{\Pubmedprefix}{pmid:}
\providecommand{\doi}[1]{\href{http://dx.doi.org/#1}{\path{#1}}}
\providecommand{\Pubmed}[1]{\href{pmid:#1}{\path{#1}}}
\providecommand{\bibinfo}[2]{#2}
\ifx\xfnm\undefined \def\xfnm[#1]{\unskip,\space#1}\fi
\bibitem[{Alam et~al.(2003)Alam, Moriya, Takai and
  Sakamoto}]{alam2003fluctuating}
\bibinfo{author}{Alam\xfnm[ M.M.]}, \bibinfo{author}{Moriya\xfnm[ M.]},
  \bibinfo{author}{Takai\xfnm[ K.]}, \bibinfo{author}{Sakamoto\xfnm[ H.]}.
\newblock \bibinfo{title}{Fluctuating fluid forces acting on two circular
  cylinders in a tandem arrangement at a subcritical reynolds number}.
\newblock \bibinfo{journal}{Journal of Wind Engineering and Industrial
  Aerodynamics}
  \bibinfo{year}{2003};\bibinfo{volume}{91}(\bibinfo{number}{1-2}):\bibinfo{pages}{139--154}.
\bibitem[{Arosio(1984)}]{arosio1984duhamel}
\bibinfo{author}{Arosio\xfnm[ A.]}.
\newblock \bibinfo{title}{Duhamel's principle for temporally inhomogeneous
  evolution equations in banach space.}
\newblock \bibinfo{journal}{NONLINEAR ANAL THEORY METHODS APPLIC}
  \bibinfo{year}{1984};\bibinfo{volume}{8}(\bibinfo{number}{9}):\bibinfo{pages}{997--1010}.
\bibitem[{Bahmani and Akbari(2010)}]{bahmani2010effects}
\bibinfo{author}{Bahmani\xfnm[ M.]}, \bibinfo{author}{Akbari\xfnm[ M.]}.
\newblock \bibinfo{title}{Effects of mass and damping ratios on viv of a
  circular cylinder}.
\newblock \bibinfo{journal}{Ocean Engineering}
  \bibinfo{year}{2010};\bibinfo{volume}{37}(\bibinfo{number}{5-6}):\bibinfo{pages}{511--519}.
\bibitem[{Deng et~al.(2020{\natexlab{a}})Deng, Ren, Xu, Fu, Moan and
  Gao}]{deng2020experimental}
\bibinfo{author}{Deng\xfnm[ S.]}, \bibinfo{author}{Ren\xfnm[ H.]},
  \bibinfo{author}{Xu\xfnm[ Y.]}, \bibinfo{author}{Fu\xfnm[ S.]},
  \bibinfo{author}{Moan\xfnm[ T.]}, \bibinfo{author}{Gao\xfnm[ Z.]}.
\newblock \bibinfo{title}{Experimental study of vortex-induced vibration of a
  twin-tube submerged floating tunnel segment model}.
\newblock \bibinfo{journal}{Journal of Fluids and Structures}
  \bibinfo{year}{2020}{\natexlab{a}};\bibinfo{volume}{94}:\bibinfo{pages}{102908}.
\bibitem[{Deng et~al.(2020{\natexlab{b}})Deng, Ren, Xu, Fu, Moan and
  Gao}]{deng2020experimentalD}
\bibinfo{author}{Deng\xfnm[ S.]}, \bibinfo{author}{Ren\xfnm[ H.]},
  \bibinfo{author}{Xu\xfnm[ Y.]}, \bibinfo{author}{Fu\xfnm[ S.]},
  \bibinfo{author}{Moan\xfnm[ T.]}, \bibinfo{author}{Gao\xfnm[ Z.]}.
\newblock \bibinfo{title}{Experimental study on the drag forces on a twin-tube
  submerged floating tunnel segment model in current}.
\newblock \bibinfo{journal}{Applied Ocean Research}
  \bibinfo{year}{2020}{\natexlab{b}};\bibinfo{volume}{104}:\bibinfo{pages}{102326}.
\bibitem[{Ehlers et~al.(2022)Ehlers, Abdussamie, Branner, Fu, Hoogeland,
  Kolari, Lara, Michailides, Murayama, Rizzo et~al.}]{ehlers2022committee}
\bibinfo{author}{Ehlers\xfnm[ S.]}, \bibinfo{author}{Abdussamie\xfnm[ N.]},
  \bibinfo{author}{Branner\xfnm[ K.]}, \bibinfo{author}{Fu\xfnm[ S.]},
  \bibinfo{author}{Hoogeland\xfnm[ M.]}, \bibinfo{author}{Kolari\xfnm[ K.]},
  \bibinfo{author}{Lara\xfnm[ P.]}, \bibinfo{author}{Michailides\xfnm[ C.]},
  \bibinfo{author}{Murayama\xfnm[ H.]}, \bibinfo{author}{Rizzo\xfnm[ C.]},
  et~al.
\newblock \bibinfo{title}{Committee v. 2: Experimental methods}.
\newblock In: \bibinfo{booktitle}{International Ship and Offshore Structures
  Congress}. \bibinfo{organization}{SNAME}; \bibinfo{year}{2022}. p.
  \bibinfo{pages}{D011S001R003}.
\bibitem[{Fu et~al.(2024)Fu, Fu, Zhang, Ren, Zhao and Xu}]{fu2024vortex}
\bibinfo{author}{Fu\xfnm[ X.]}, \bibinfo{author}{Fu\xfnm[ S.]},
  \bibinfo{author}{Zhang\xfnm[ M.]}, \bibinfo{author}{Ren\xfnm[ H.]},
  \bibinfo{author}{Zhao\xfnm[ B.]}, \bibinfo{author}{Xu\xfnm[ Y.]}.
\newblock \bibinfo{title}{Vortex-induced vibration of a flexible pipe under
  oscillatory sheared flow}.
\newblock \bibinfo{journal}{Physical Review Fluids}
  \bibinfo{year}{2024};\bibinfo{volume}{9}(\bibinfo{number}{1}):\bibinfo{pages}{014604}.
\bibitem[{Govardhan and Williamson(2000)}]{govardhan2000modes}
\bibinfo{author}{Govardhan\xfnm[ R.]}, \bibinfo{author}{Williamson\xfnm[ C.]}.
\newblock \bibinfo{title}{Modes of vortex formation and frequency response of a
  freely vibrating cylinder}.
\newblock \bibinfo{journal}{Journal of Fluid Mechanics}
  \bibinfo{year}{2000};\bibinfo{volume}{420}:\bibinfo{pages}{85--130}.
\bibitem[{Govardhan and Williamson(2002)}]{govardhan2002resonance}
\bibinfo{author}{Govardhan\xfnm[ R.]}, \bibinfo{author}{Williamson\xfnm[ C.]}.
\newblock \bibinfo{title}{Resonance forever: existence of a critical mass and
  an infinite regime of resonance in vortex-induced vibration}.
\newblock \bibinfo{journal}{Journal of Fluid Mechanics}
  \bibinfo{year}{2002};\bibinfo{volume}{473}:\bibinfo{pages}{147--166}.
\bibitem[{Govardhan and Williamson(2006)}]{govardhan2006defining}
\bibinfo{author}{Govardhan\xfnm[ R.]}, \bibinfo{author}{Williamson\xfnm[ C.]}.
\newblock \bibinfo{title}{Defining the ‘modified griffin plot’in
  vortex-induced vibration: revealing the effect of reynolds number using
  controlled damping}.
\newblock \bibinfo{journal}{Journal of fluid mechanics}
  \bibinfo{year}{2006};\bibinfo{volume}{561}:\bibinfo{pages}{147--180}.
\bibitem[{Hover et~al.(1998)Hover, Techet and Triantafyllou}]{hover1998forces}
\bibinfo{author}{Hover\xfnm[ F.]}, \bibinfo{author}{Techet\xfnm[ A.]},
  \bibinfo{author}{Triantafyllou\xfnm[ M.]}.
\newblock \bibinfo{title}{Forces on oscillating uniform and tapered cylinders
  in cross flow}.
\newblock \bibinfo{journal}{Journal of Fluid Mechanics}
  \bibinfo{year}{1998};\bibinfo{volume}{363}:\bibinfo{pages}{97--114}.
\bibitem[{Janocha et~al.(2021)Janocha, Ong, Nystr{\o}m, Tu, Endal and
  Stokholm}]{janocha2021flow}
\bibinfo{author}{Janocha\xfnm[ M.J.]}, \bibinfo{author}{Ong\xfnm[ M.C.]},
  \bibinfo{author}{Nystr{\o}m\xfnm[ P.R.]}, \bibinfo{author}{Tu\xfnm[ Z.]},
  \bibinfo{author}{Endal\xfnm[ G.]}, \bibinfo{author}{Stokholm\xfnm[ H.]}.
\newblock \bibinfo{title}{Flow around two elastically-mounted cylinders with
  different diameters in tandem and staggered configurations in the subcritical
  reynolds number regime}.
\newblock \bibinfo{journal}{Marine Structures}
  \bibinfo{year}{2021};\bibinfo{volume}{76}:\bibinfo{pages}{102893}.
\bibitem[{Khalak and Williamson(1997)}]{khalak1997investigation}
\bibinfo{author}{Khalak\xfnm[ A.]}, \bibinfo{author}{Williamson\xfnm[ C.H.]}.
\newblock \bibinfo{title}{Investigation of relative effects of mass and damping
  in vortex-induced vibration of a circular cylinder}.
\newblock \bibinfo{journal}{Journal of Wind Engineering and Industrial
  Aerodynamics}
  \bibinfo{year}{1997};\bibinfo{volume}{69}:\bibinfo{pages}{341--350}.
\bibitem[{Mackowski and Williamson(2011)}]{mackowski2011developing}
\bibinfo{author}{Mackowski\xfnm[ A.W.]}, \bibinfo{author}{Williamson\xfnm[
  C.H.]}.
\newblock \bibinfo{title}{Developing a cyber-physical fluid dynamics facility
  for fluid--structure interaction studies}.
\newblock \bibinfo{journal}{Journal of Fluids and Structures}
  \bibinfo{year}{2011};\bibinfo{volume}{27}(\bibinfo{number}{5-6}):\bibinfo{pages}{748--757}.
\bibitem[{Mysa et~al.(2016)Mysa, Kaboudian and Jaiman}]{mysa2016origin}
\bibinfo{author}{Mysa\xfnm[ R.C.]}, \bibinfo{author}{Kaboudian\xfnm[ A.]},
  \bibinfo{author}{Jaiman\xfnm[ R.K.]}.
\newblock \bibinfo{title}{On the origin of wake-induced vibration in two tandem
  circular cylinders at low reynolds number}.
\newblock \bibinfo{journal}{Journal of Fluids and Structures}
  \bibinfo{year}{2016};\bibinfo{volume}{61}:\bibinfo{pages}{76--98}.
\bibitem[{Ren et~al.(2024)Ren, Fu, Zhang, Xu and Ren}]{ren2024developing}
\bibinfo{author}{Ren\xfnm[ H.]}, \bibinfo{author}{Fu\xfnm[ S.]},
  \bibinfo{author}{Zhang\xfnm[ M.]}, \bibinfo{author}{Xu\xfnm[ Y.]},
  \bibinfo{author}{Ren\xfnm[ H.]}.
\newblock \bibinfo{title}{Developing a virtual physical system for
  vortex-induced vibration studies of a bluff body}.
\newblock \bibinfo{journal}{Journal of Ocean Engineering and Science}
  \bibinfo{year}{2024};.
\bibitem[{Shen et~al.(2024)Shen, Fu, Zhang, Niu, Hua, Xu, Chu and
  Bin}]{shen2024experimental}
\bibinfo{author}{Shen\xfnm[ J.]}, \bibinfo{author}{Fu\xfnm[ S.]},
  \bibinfo{author}{Zhang\xfnm[ M.]}, \bibinfo{author}{Niu\xfnm[ Z.]},
  \bibinfo{author}{Hua\xfnm[ Y.]}, \bibinfo{author}{Xu\xfnm[ Y.]},
  \bibinfo{author}{Chu\xfnm[ Y.]}, \bibinfo{author}{Bin\xfnm[ S.]}.
\newblock \bibinfo{title}{Experimental investigation on vortex induced
  vibration of a rigidly coupled twin-tube model}.
\newblock In: \bibinfo{booktitle}{International Conference on Offshore
  Mechanics and Arctic Engineering}. \bibinfo{organization}{American Society of
  Mechanical Engineers}; volume \bibinfo{volume}{87844}; \bibinfo{year}{2024}.
  p. \bibinfo{pages}{V006T08A018}.
\bibitem[{Song et~al.(2016)Song, Fu, Cao, Ma and Wu}]{song2016investigation}
\bibinfo{author}{Song\xfnm[ L.]}, \bibinfo{author}{Fu\xfnm[ S.]},
  \bibinfo{author}{Cao\xfnm[ J.]}, \bibinfo{author}{Ma\xfnm[ L.]},
  \bibinfo{author}{Wu\xfnm[ J.]}.
\newblock \bibinfo{title}{An investigation into the hydrodynamics of a flexible
  riser undergoing vortex-induced vibration}.
\newblock \bibinfo{journal}{Journal of Fluids and Structures}
  \bibinfo{year}{2016};\bibinfo{volume}{63}:\bibinfo{pages}{325--350}.
\bibitem[{Vandiver(1983)}]{vandiver1983drag}
\bibinfo{author}{Vandiver\xfnm[ J.K.]}.
\newblock \bibinfo{title}{Drag coefficients of long flexible cylinders}.
\newblock In: \bibinfo{booktitle}{Offshore technology conference}.
  \bibinfo{organization}{OTC}; \bibinfo{year}{1983}. p.
  \bibinfo{pages}{OTC--4490}.
\bibitem[{Williamson and Govardhan(2008)}]{williamson2008brief}
\bibinfo{author}{Williamson\xfnm[ C.]}, \bibinfo{author}{Govardhan\xfnm[ R.]}.
\newblock \bibinfo{title}{A brief review of recent results in vortex-induced
  vibrations}.
\newblock \bibinfo{journal}{Journal of Wind engineering and industrial
  Aerodynamics}
  \bibinfo{year}{2008};\bibinfo{volume}{96}(\bibinfo{number}{6-7}):\bibinfo{pages}{713--735}.
\bibitem[{Williamson and Govardhan(2004)}]{williamson2004vortex}
\bibinfo{author}{Williamson\xfnm[ C.H.]}, \bibinfo{author}{Govardhan\xfnm[
  R.]}.
\newblock \bibinfo{title}{Vortex-induced vibrations}.
\newblock \bibinfo{journal}{Annu Rev Fluid Mech}
  \bibinfo{year}{2004};\bibinfo{volume}{36}(\bibinfo{number}{1}):\bibinfo{pages}{413--455}.
\bibitem[{Zhang and Haque(2022)}]{zhang2022wake}
\bibinfo{author}{Zhang\xfnm[ K.]}, \bibinfo{author}{Haque\xfnm[ M.N.]}.
\newblock \bibinfo{title}{Wake interactions between two side-by-side circular
  cylinders with different sizes}.
\newblock \bibinfo{journal}{Physical Review Fluids}
  \bibinfo{year}{2022};\bibinfo{volume}{7}(\bibinfo{number}{6}):\bibinfo{pages}{064703}.
\bibitem[{Zhao et~al.(2023{\natexlab{a}})Zhao, Zhang, Fu, Fu, Ren and
  Xu}]{zhao2023drag}
\bibinfo{author}{Zhao\xfnm[ B.]}, \bibinfo{author}{Zhang\xfnm[ M.]},
  \bibinfo{author}{Fu\xfnm[ S.]}, \bibinfo{author}{Fu\xfnm[ X.]},
  \bibinfo{author}{Ren\xfnm[ H.]}, \bibinfo{author}{Xu\xfnm[ Y.]}.
\newblock \bibinfo{title}{Drag coefficients of double unequal-diameter flexible
  cylinders in tandem undergoing vortex/wake-induced vibrations}.
\newblock \bibinfo{journal}{Ocean Engineering}
  \bibinfo{year}{2023}{\natexlab{a}};\bibinfo{volume}{270}:\bibinfo{pages}{113642}.
\bibitem[{Zhao et~al.(2023{\natexlab{b}})Zhao, Zhang, Fu, Fu, Sun and
  Song}]{zhao2023experimental}
\bibinfo{author}{Zhao\xfnm[ B.]}, \bibinfo{author}{Zhang\xfnm[ M.]},
  \bibinfo{author}{Fu\xfnm[ S.]}, \bibinfo{author}{Fu\xfnm[ X.]},
  \bibinfo{author}{Sun\xfnm[ T.]}, \bibinfo{author}{Song\xfnm[ B.]}.
\newblock \bibinfo{title}{Experimental investigation on vortex/wake-induced
  force of double unequal-diameter flexible cylinders in tandem}.
\newblock \bibinfo{journal}{Physics of Fluids}
  \bibinfo{year}{2023}{\natexlab{b}};\bibinfo{volume}{35}(\bibinfo{number}{5}).
\bibitem[{Zhao(2013)}]{zhao2013flow}
\bibinfo{author}{Zhao\xfnm[ M.]}.
\newblock \bibinfo{title}{Flow induced vibration of two rigidly coupled
  circular cylinders in tandem and side-by-side arrangements at a low reynolds
  number of 150}.
\newblock \bibinfo{journal}{Physics of Fluids}
  \bibinfo{year}{2013};\bibinfo{volume}{25}(\bibinfo{number}{12}).
\bibitem[{Zhu et~al.(2023)Zhu, Zhao, Qiu, Lin, Du and Dong}]{zhu2023vortex}
\bibinfo{author}{Zhu\xfnm[ H.]}, \bibinfo{author}{Zhao\xfnm[ Y.]},
  \bibinfo{author}{Qiu\xfnm[ T.]}, \bibinfo{author}{Lin\xfnm[ W.]},
  \bibinfo{author}{Du\xfnm[ X.]}, \bibinfo{author}{Dong\xfnm[ H.]}.
\newblock \bibinfo{title}{Vortex-induced vibrations of two tandem rigidly
  coupled circular cylinders with streamwise, transverse, and rotational
  degrees of freedom}.
\newblock \bibinfo{journal}{Physics of Fluids}
  \bibinfo{year}{2023};\bibinfo{volume}{35}(\bibinfo{number}{2}).

\end{thebibliography}

\end{document}